\documentclass[runningheads]{llncs}
\usepackage{graphicx}
\usepackage[utf8]{inputenc} 
\usepackage[T1]{fontenc}    
\usepackage{xr-hyper}
\usepackage{xcolor}
\usepackage{hyperref}       
\usepackage{url}            
\usepackage{booktabs}       
\usepackage{amsfonts}       
\usepackage{amsmath}       
\usepackage{nicefrac}       
\usepackage{microtype}      
\usepackage{epsfig}
\usepackage{array,multirow}
\usepackage{multicol}
\usepackage{xspace}
\usepackage{cleveref}[capitalise]

\usepackage{adjustbox}
\usepackage{diagbox}
\usepackage{floatrow}
\usepackage[font=small,labelfont=bf]{caption}
\newfloatcommand{capbtabbox}{table}[][\FBwidth]

\makeatletter
\DeclareRobustCommand\onedot{\futurelet\@let@token\@onedot}
\def\@onedot{\ifx\@let@token.\else.\null\fi\xspace}

\def\eg{\emph{e.g}\onedot}

\def\etal{\emph{et al}\onedot}
\makeatother

\graphicspath{{./fig/}}

\begin{document}
\title{Metadata Improves Segmentation Through Multitasking Elicitation}
%
%
\author{
Iaroslav Plutenko\inst{1,2} \and
Mikhail Papkov\inst{3} \and
Kaupo Palo\inst{4} \and
Leopold Parts\inst{3,5} \and
Dmytro Fishman\inst{3}
}
\authorrunning{I. Plutenko et al.}
%
\institute{
Ukrainian Catholic University, Lviv, Ukraine \and 
Leibniz Institute of Plant Genetics and Crop Plant Research (IPK), Gatersleben, Germany \and 
Institute of Computer Science, University of Tartu, Estonia \and
Revvity, Inc., Tallinn, Estonia \and
Wellcome Sanger Institute, Hinxton, United Kingdom  \\
\email{plutenko@ipk-gatersleben.de \hspace{2cm} dmytro.fishman@ut.ee}
}
%

%
\maketitle              
\begin{abstract}

Metainformation is a common companion to biomedical images. However, this potentially powerful additional source of signal from image acquisition has had limited use in deep learning methods, for semantic segmentation in particular. Here, we incorporate metadata by employing a channel modulation mechanism in convolutional networks and study its effect on semantic segmentation tasks. We demonstrate that metadata as additional input to a convolutional network can improve segmentation results while being inexpensive in implementation as a nimble add-on to popular models. We hypothesize that this benefit of metadata can be attributed to facilitating multitask switching. This aspect of metadata-driven systems is explored and discussed in detail.

\end{abstract}
\section{Introduction}
Semantic segmentation is a technique broadly applied in biomedicine, separating critical regions according to their functional role in the source object, \eg, detecting tumors from CT images~\cite{Litjens2017-ah} or differentiating cell nuclei from the background~\cite{Ali2021-mi}. When performed by trained human experts, this task is time-consuming and costly~\cite{Bhalgat2018-hl}. The oldest automated techniques~\cite{Otsu1979-kn,Beucher1979-fs} in image processing were rule-based routines that separated regions on deterministic criteria. Since then, machine learning methods have taken over the field~\cite{Georgiou2020-yh,Khan2021-yf}. Most recently, deep learning models eliminated many manual steps in image processing and outperformed previous methods~\cite{Greenspan2016-cz}, but substantial improvements are still needed for robust systems.

Digital images often come with abundant metainformation reflecting image acquisition device settings, the methodology of sample preparation, or the provenance of the inspected objects. This information alone can say a lot about how images look and their quality, and potentially influence the segmentation results. For example, different cell lines have distinct appearances under the microscope~\cite{Ali2021-mi}, and knowing which types of cells to look for helps locate them.

Incorporating metadata into computer vision models is straightforward for the classification task. Kawahara~\etal used a convolutional neural network (CNN) to extract features from skin images and concatenated them with a one-hot encoded metadata vector to analyze lesions~\cite{Kawahara2018-yq}. Gessert~\etal used an additional dense neural network to process age, anatomical site, and sex metadata and fused its features with the CNN~\cite{Gessert2020-qp}. However, mixing non-image data into the fully-convolutional neural network for semantic segmentation is less trivial. 
De Vries~\etal introduced Conditional Batch Normalization to integrate additional information into feature maps~\cite{De_Vries2017-du}. This idea was developed into Feature-wise Linear Modulation (FiLM)~\cite{Perez2018-aw} and Conditioned-U-Net~\cite{Brocal2019-at}. Later, FiLM was adopted for the biomedical domain~\cite{Lemay2021-jn}.

The idea of modulating CNN feature maps closely relates to channel attention, which uses learnable weights to emphasize relevant feature maps. Channel attention is actively employed to improve CNN's performance~\cite{Woo2018-gk}. Hu~\etal introduced a squeeze-and-excitation (SE) block to implement this idea by channel-wise feature recalibration~\cite{Hu2020-su}. Lightweight channel modulation blocks were beneficial for CNNs in various tasks~\cite{Roy2020-vm,Amer2020-mg}. 

This work explores the value of channel attention using SE block for metadata incorporation. We investigate the effects of categorical and continuous metadata on semantic segmentation and show that metadata improves the model's generalization and overall performance. We conduct experiments on biomedical datasets using metadata such as cell line labels and expected object size. Further, we explore the utility of the metadata to help the model effectively navigate multiple tasks on the same images. The results show a statistically significant advantage of metadata-driven models over metadata-free models. The difference becomes more pronounced when the model is trained on visually similar domains, and the system exhibits its multitasking properties. Overall, our contributions are as follows:
\begin{enumerate}
    \item We present a novel, simple, lightweight, yet effective metadata-enhanced squeeze-and-excitation block. 
    \item We empirically show that using metadata improves the performance of the semantic segmentation models.
    \item We find that metadata drastically increases the performance for the underrepresented task in a multitask setting.
\end{enumerate}

\section{Methods}

\subsection{Metadata incorporation with channel modulation}
\label{sec:methods}
Squeeze-and-excitation (SE) blocks are a natural place for metadata fusion in a shallow bottleneck for channel recalibration. They consist of two linear layers, with ReLU activation for the first layer and sigmoid activation for the second. Each block's input is squeezed from feature maps with pooling. The output is multiplied channel-wise with the same feature maps. We propose two ways of modifying an SE block. First, encoded metainformation can fully replace the squeezed input to the multi-layer perceptron. We call this block metadata-excitation (ME) because we abandon squeezing entirely. Second, we can concatenate metainformation to the squeezed vector allowing the network itself to decide on the importance of each input. We refer to this modification as squeeze-metadata-and-excitation (SME). Here, we test the performance of ME and SME blocks as components of U-Net network architecture against the baseline model with vanilla SE block.

\begin{figure}[h]
\begin{center}
\includegraphics[width=0.97\linewidth]{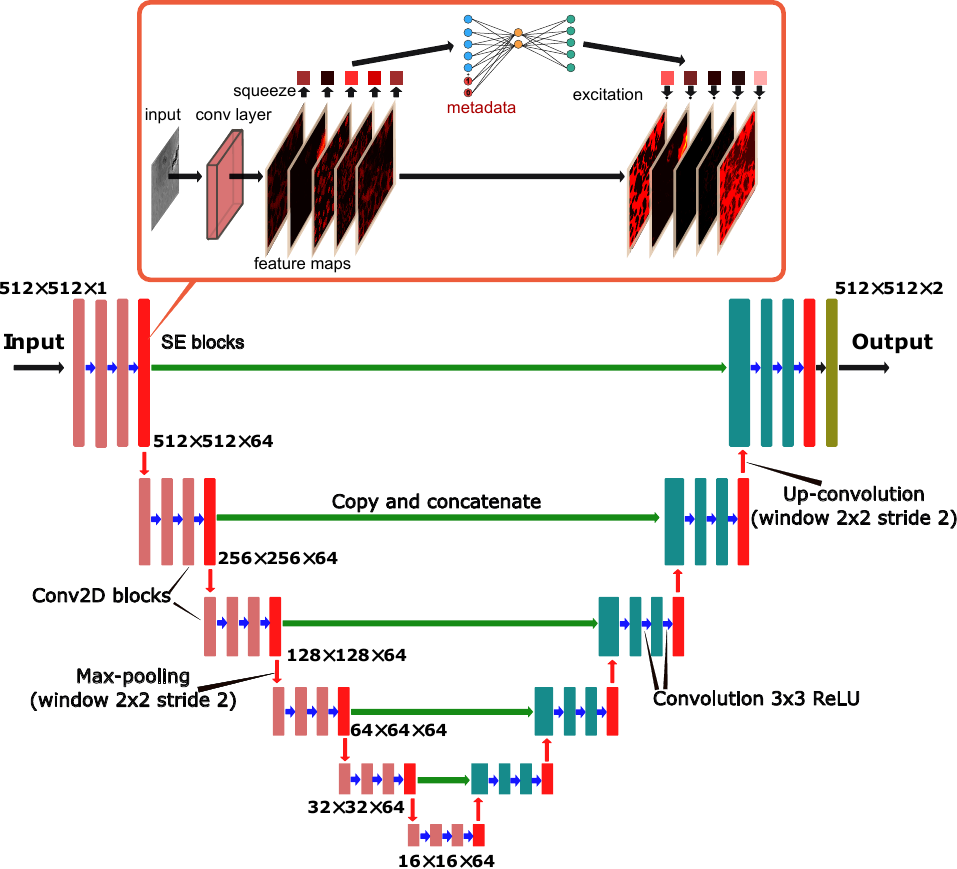}
\end{center}
   \caption{Integration of metadata into U-Net as an additional block with channel attention functionality that can also receive encoded metadata affecting feature maps intensity. Here the schema for the SME model is shown. For the ME model squeeze part is absent, and the input to the linear layers consists of metadata only (red circles). }
\label{fig:unet}
\end{figure}

We implement ME and SME blocks as a part of the modified U-Net~\cite{Ronneberger2015-me} architecture (\Cref{fig:unet}). The number of inputs depends on the configuration (number of feature maps, metadata elements, and selected type of block). Hidden dimensionality is four times lower than the number of feature maps.
\subsection{Training}
We used Adam optimizer \cite{Kingma2014-hg} with CyclicLR~\cite{Smith2017-sf} scheduler changing the learning rate linearly after each batch from $0.0002$ to $0.0008$ and back in 8 cycles for 100 epochs. We saved the best model by validation error for the downstream analysis. We implemented all the models in PyTorch 1.6~\cite{paszke2019pytorch} and trained them on NVIDIA Tesla V100 16/32Gb GPUs with cuDNN 7.6.3.

 \section{Experiments}

\subsection{Microscopy segmentation across cell lines}
To assess the impact of metadata on microscopy image segmentation, we trained our models on the Seven Cell Lines brightfield microscopy dataset~\cite{Ali2021-mi}. It consists of 3024 samples from seven cell lines in equal proportions: HeLa, MDCK, HepG2, A549, HT1080, MCF7, NIH3T3. Each sample is a single-channel image of size 1080 x 1080 pixels paired with a binary mask of cell nuclei and a metadata vector with a one-hot encoded cell line label (Supplementary~\Cref{table:seven_lines_distr}). Part of the models was trained conventionally without access to metadata, and another part had been assisted by metadata labels provided as additional input. Our best models that had access to metadata (SME) had shown an improvement in $F_1$ score from 0.861 to 0.863 (p < 0.001) compared to the baseline model trained without metainformation (\Cref{table:seven_lines_scores}). For comparison, we also used the FiLM model from the \texttt{ivadomed} toolbox~\cite{Gros2021-nm}. It did not outperform the baseline model with an F1 score of 0.86. Visually the improvements are seen as better overlaps with ground truth masks (Supplementary \Cref{fig:sup_results-7cl}).

\begin{table}[ht]
\centering

\begin{tabular}{lcccccccc}
\toprule
& Average & HeLa & MDCK & A549 & HT1080 & HepG2 & MCF7 & NIH3T3 \\

\midrule
baseline stratified & -                           & \underline{0.901}                    & 0.830                     & 0.841                    & 0.857                      & 0.806                     & 0.825                    & \underline{0.891}                      \\
baseline                                                      & \underline{0.861}                       & \underline{0.901}                    & \underline{0.865}                    & 0.852                    & \underline{0.870}                       & 0.806                     & \underline{0.840}                     & \underline{0.891}                      \\
FiLM-dummy~\cite{Perez2018-aw,Brocal2019-at,Lemay2021-jn}                                                     & 0.855                       & 0.895                    & 0.858                    & 0.846                    & 0.864                      & 0.802                     & 0.835                    & 0.885                      \\
FiLM~\cite{Perez2018-aw,Brocal2019-at,Lemay2021-jn}                                                           & 0.860                        & 0.900                      & \underline{0.865}                    & \underline{0.853}                    & 0.867                      & 0.806                     & 0.839                    & 0.890                       \\
ME                                                            & \underline{0.861}                       & 0.900                      & \underline{0.865}                    & 0.852                    & 0.869                      & \underline{0.808}                     & \textbf{0.842}                    & \underline{0.891}                      \\
SME-dummy                                                     & 0.859                       & 0.899                    & 0.864                    & 0.852                    & 0.867                      & 0.805                     & 0.838                    & 0.889                      \\
SME                                                           & \textbf{0.863}                       & \textbf{0.902}                    & \textbf{0.869}                    & \textbf{0.854}                    & \textbf{0.872}                      & \textbf{0.809}                     & \textbf{0.842}                    & \textbf{0.893}                      
\\
\bottomrule
\end{tabular}

\caption{Resulting $F_1$ scores of models trained on the Seven Cell Lines dataset. See \Cref{sec:methods} for method description. ``baseline stratified'' denotes the performance of the models trained on the particular subset of data (cell line),  ``dummy'' denotes metadata vector filled with zeros. The best score is in \textbf{bold}, the second best \underline{underlined}.
}
\label{table:seven_lines_scores}
\end{table}

\subsection{Microscopy segmentation across annotation styles}

Microscopists can exhibit different annotation styles when drawing an object boundary as polygon vertices. More points provide more accurate segmentation, while fewer points save labeling time. From the Seven Cell Lines training dataset we derived a dataset with three nuclei stencil styles using various polygonization degrees of \texttt{skimage.measure.approximate\_polygon()} ~\cite{van2014scikit}. One-third of masks were left unchanged. The second third had a fine approximation to polygons. The remaining part had coarser annotations (Supplementary~\Cref{fig:sup_altered-polygons}). The test and validation parts had accurate masks and remained consistent with the Seven Cell Lines dataset. Thus, we aimed to obtain fine-grained segmentation results with coarsely annotated parts of the training set.

Here, a metadata label serves as a task identifier, making the model predict masks with shape and confidence specific to the annotation style. The coarse style lowers the prediction confidence, affecting the area of the binarized mask. Switching the model to the mode with accurate masks during the evaluation boosts confidence and helps increase accuracy on the test set (Supplementary \Cref{fig:sup_results-polygons}).
All models with metadata had higher $F_1$ scores, with improvements from 0.846 to 0.854 with our best SME model compared to the baseline model trained on all three parts of the dataset without metadata ($p < 0.001$, \Cref{table:polygon_dataset}).

\begin{table}[ht]
\centering
\begin{tabular}{lc}
\toprule
          & $F_1$ score \\
\midrule          
baseline 1/3 data    & 0.837                                 \\
baseline             & 0.846                                 \\
ME                  & \underline{0.850}                                 \\
SME               & \textbf{0.854}                                 \\
FiLM                 & 0.849                                 \\
SME dummy            & 0.848                                                     
\\ \bottomrule
\end{tabular}
\caption{Results of the experiment with the dataset with different annotation styles. Row ``baseline 1/3'' denotes results from the SE model trained on one-third of data with accurate masks. The main baseline model was trained on the full dataset without metadata. ME, SME, and FiLM models use metadata with different implementations. SME dummy - SME model with meaningless metadata (zeros in the input vector). The best score is in \textbf{bold}, the second best \underline{underlined}.
}
\label{table:polygon_dataset}
\end{table}

\subsection{Microscopy multilabel segmentation: cells and anomalies}

Researchers prefer a clean dataset, as visual artifacts hinder downstream analysis~\cite{Ali2022-ry}. However, anomaly segmentation for microscopy remains a challenging task because of the lack of annotations and their diverse shapes and sizes. We propose to take advantage of nuclei mask abundance and use metadata as a switch for multitask learning. Thus, we train the network to produce nuclei and anomaly segmentation from a single head, depending on the metadata input.  To assess this approach and compare it with a conventional parallel multi-headed prediction, we expand the Seven Cell Lines dataset with multilabel targets.  
For 365 images out of 3024, we additionally present anomaly segmentation masks, highlighting debris and optical defects. For these images, metadata encodes the mask type: nuclei or anomaly. Since anomaly annotations are absent for the rest of the dataset, we are able to train parallel multi-headed networks only on this subset with two segmentation masks. On the contrary, we can train on the whole dataset when using metadata as a task switch.

A metadata-driven model achieved an $F_1$ score of 0.836 and 0.85 for nuclei and anomalies, respectively (\Cref{table:anom_dataset}). It notably outperformed the baseline model trained to segment only anomalies that had $F_1$ score of 0.736 on the anomaly subset. At the same time, the performance for nuclei segmentation slightly dropped from the baseline $F_1$ score of 0.854 on the nuclei subset. 
Expectedly, swapping metadata labels resulted in segmentation failure, approaching zero F1 score. The model with dummy metadata input also struggled with anomaly segmentation due to the dominance of nuclei masks, predicting only nuclei.

Additionally, we utilized a limited version of the large dataset containing nuclei and anomalies on the same images. Training a multi-headed model parallelly on this reduced dataset, with each head's output corresponding to a different mask, yielded a higher $F_1$ score of 0.835 on the anomaly subset than the baseline model with $F_1$ score of 0.736 training exclusively on that subset. The performance on the nuclei mask subset was lower due to the limited training size (the stratified baseline model trained on a limited subset of nuclei was not introduced in this experiment due to the focus on anomaly segmentation).

Surprisingly, when our initial metadata-driven model with a single output was retrained on the reduced dataset in the usual sequential mode using metadata to switch segmentation task, the anomaly subset showed the highest boost in $F_1$ score among all experiments: 0.854 compared to the individual model with $F_1$ score of 0.736. Detailed results are summarized in~\Cref{table:anom_dataset}.

\begin{table}[ht]
\centering
\begin{tabular}{llccc}
\toprule
 & & {Average}     & Anomalies & Seven Cell Lines \\ \midrule
\multirow{6}{*}{{Full dataset}} & {baseline stratified}                                     & - & 0.736              & 0.854                                                                \\
&{ME}                                                      & 0.835                & {0.824}     & 0.845                                                                \\
&{SME-dummy}                                               & 0.470                & 0.104              & 0.835                                                               
\\
&{SME}                                                     & 0.843                & \underline{0.836}     & 0.850                                                                \\
\midrule                                                       
\multirow{2}{*}{{Multilabel subset}} &{Two heads}                                                               & 0.812                & 0.835                          & 0.789                                 \\
&{SME}                                                                   & 0.823                &  \textbf{0.854}                 & 0.792                                
\\ \bottomrule
\end{tabular}
\caption{$F_1$ scores for multitask models from experiments on dataset with anomalies (full and multilabel subset). See \Cref{sec:methods} for method description. ``baseline stratified'' denotes the performance of the SE models trained on the particular subset of data (anomaly/nuclei masks), ``dummy'' denotes the model with metadata vector filled with zeros. The best score is in \textbf{bold}, the second best \underline{underlined}.
}
\label{table:anom_dataset}
\end{table}

\begin{figure}[h]
\begin{center}
\includegraphics[width=0.99\linewidth]{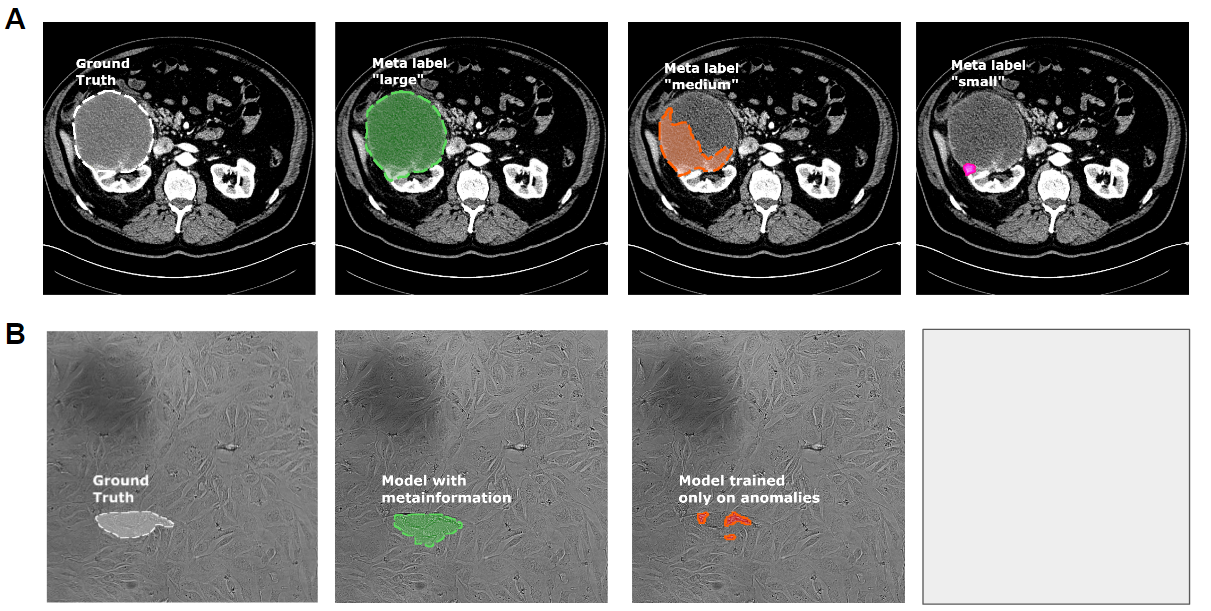}
\end{center}
   \caption{(A) Manifestation of multitask properties in metadata-driven models with continuous metadata correlated with the tumor size on KiTS21 dataset. In this experiment, the baseline model had on the test $F_1$ score of 0.734, and metadata-driven models (SME and ME) had an average $F_1$ score of 0.804 for two models. (B) Example of positive transfer for the anomaly segmentation task with the metadata-driven model~--- the baseline model (trained only on the anomaly subset) had $F_1$ score of 0.736 on this subset, metadata-driven models (SME) had $F_1$ score of 0.854 on anomaly subset.
}
\label{fig:results-kits}
\end{figure}

\subsection{Kidney tumor segmentation with continuous metadata}
To test our method's applicability for continuous metadata, we used KiTS21~\cite{Heller2021-rc} 3D dataset of 200 images accompanied with the tumor size information. We derived a 2D dataset from it with 10346 images selecting axial slices containing kidneys. Segmentation targets contain three classes: kidney tissue, tumor, and cyst (cyst and tumors were exclusive classes~--- cases with tumors did not have cysts and vice versa). In this experiment, the baseline model had a $F_1$ score of 0.734 on the test, metadata-driven model ME achieved a $F_1$ score of 0.807, and another metadata-driven SME model achieved a $F_1$ score of 0.801. 
In~\Cref{fig:results-kits}A, we illustrate how the target mask changes in response to different size-correlated metadata applied to the same image. 

\section{Discussion}
Metadata is a widely available and potentially valuable source of additional information for deep learning systems. We used a channel modulation mechanism to inject it into the image-processing workflow, a lightweight and low-intrusive approach for adapting existing CNNs. Despite small dimensionality compared to the main input, metadata can drastically change the course of learning and inference by indicating the presence of a specific task. Our experiments show how this change is especially obvious when no other visual clues are present in the image. The results demonstrate that the CNN models endowed with metadata awareness respond strongly to the metadata input, adjusting their output according to the respective task indicator.  While the correct metadata labels improve the segmentation performance, providing incorrect labels significantly hinders the model's functionality. The metadata-driven system inherently has a multitasking nature that allows exploiting the benefits of combining several tasks. Our results for segmenting anomalies and nuclei from microscopy images confirm these assumptions.

We have proposed a method of incorporating metadata from a variety of sources into an image segmentation pipeline and demonstrated its effectiveness. We showed that metadata can act as a guiding label (in both discrete and continuous form) or even as a task switch. We believe that principles of multitask learning, some of which we have uncovered here, will remain universal for metadata-driven segmentation systems and persist regardless of chosen implementation and architecture. Ultimately, metadata could be the unifying modality that allows training truly cross-domain models.

\subsubsection{Acknowledgements}

This work was funded by Revvity, Inc. (previously known as PerkinElmer Inc., VLTAT19682), and Wellcome Trust (206194). We thank High Performance Computing Center of the Institute of Computer Science at the University of Tartu for the provided computing power.

%
%
%
\bibliographystyle{splncs04}
\bibliography{ref}


\clearpage
\setcounter{page}{1}
\renewcommand{\thetable}{S\arabic{table}}
\renewcommand{\thefigure}{S\arabic{figure}}
\setcounter{figure}{0}
\setcounter{table}{0}

\section*{Supplementary materials}

\begin{figure}[h]
\begin{center}
\includegraphics[width=0.90\linewidth]{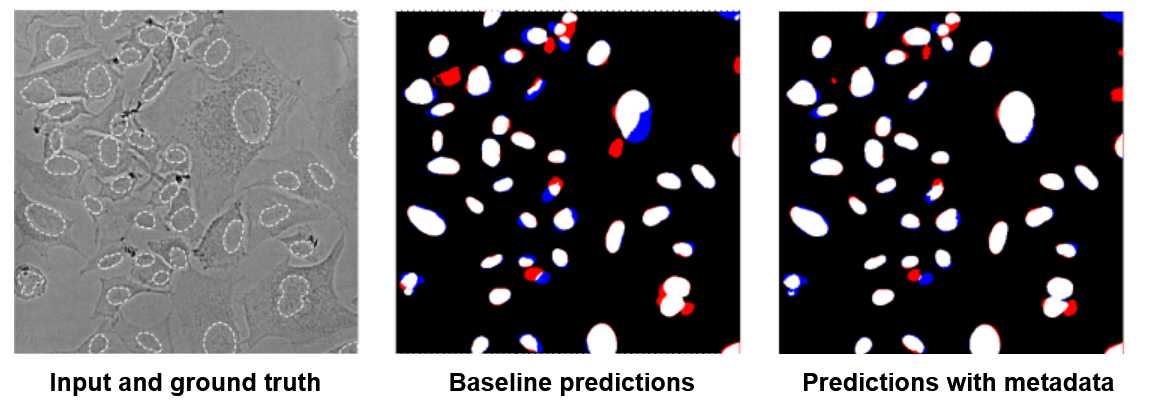}
\end{center}
   \caption{Models with metadata perform better. An example of a prediction of the A549 sample from the original Seven Cell Lines dataset mapped to the ground truth mask, where pixel color represents {\color{red} false positive}, {\color{blue} false negative}, and white denotes the correct match with ground truth. The baseline model has an average $F_1$ score of 0.861, metadata-driven SME model has an average $F_1$ score of 0.863} 
\label{fig:sup_results-7cl}
\end{figure}

\begin{figure}[h]
\begin{center}
\includegraphics[width=0.7\linewidth]{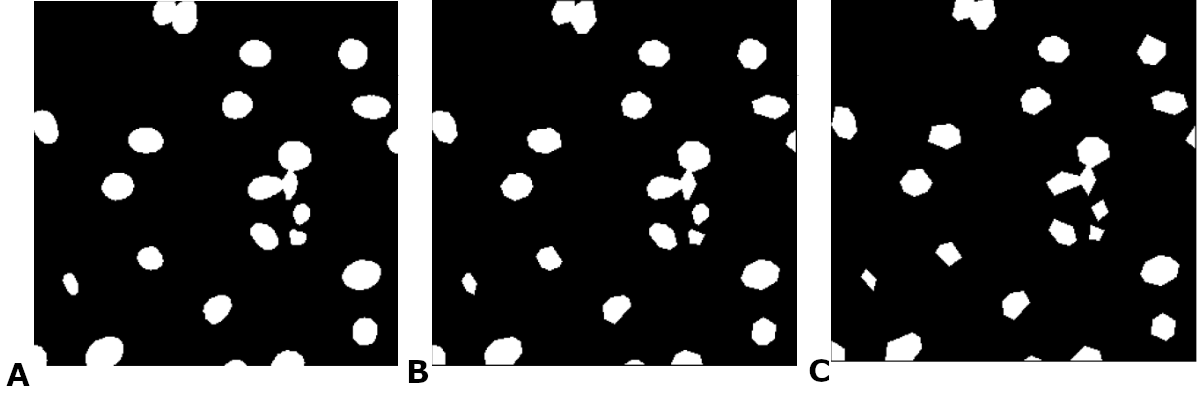}
\end{center}
  \caption{Simplified poligonized masks from cell lines dataset to imitate annotation style. Original masks (A) were poligonized with high approximation using the \texttt{tolerance=2} in \texttt{skimage.measure.approximate\_polygon()} (B) and lower approximation with the \texttt{tolerance=3.5}(C).}
\label{fig:sup_altered-polygons}
\end{figure}

\begin{figure}[h]
\centering
\includegraphics[width=0.7\linewidth]{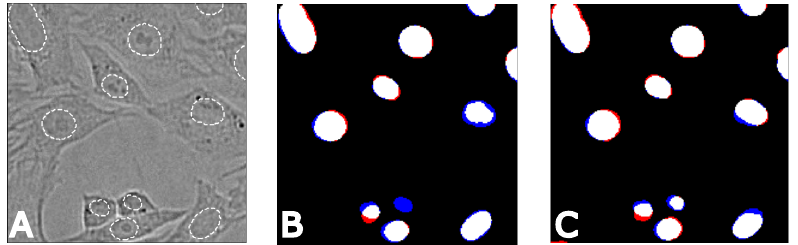}

   \caption{Prediction with metadata corresponding to accurate masks improves segmentation with more accurate stencils shape and confidence. Example of performance on a sample patch from the test data. Source image with ground truth contours (A), masks from baseline model (B), which has a test set $F_1$ score of 0.846, and masks from the SME model in the inference mode with metadata corresponding to accurate masks (C), having a test set $F_1$ score of 0.854. The color represents {\color{red} false positive} and {\color{blue} false negative} pixels and the white color denotes the correct match with ground truth.}
\label{fig:sup_results-polygons}
\end{figure}

\begin{table}[]
\begin{tabular}{lccccccccc}
\toprule

  & A549 & HT1080 & HeLa    & HepG2 & MCF7 & MDCK    & NIH3T3 & Polygons & Anomalies \\ \midrule
Train   & 286  & 284    & 293 & 283   & 290  & 292 & 288    & 672/672/672   & 194       \\
Val & 66   & 78     & 58   & 82    & 70   & 79   & 71     & 504    & 70        \\
Test       & 80   & 70     & 81   & 67    & 72   & 61   & 73     & 504    & 101  \\
\bottomrule
\end{tabular}
\caption{Datasets distribution. For the dataset with polygonized masks the distribution between accurate masks, fine and coarse polygons in the train set are shown, while the validation and test sets contain accurate masks}
\label{table:seven_lines_distr}
\end{table}

\end{document}